\documentclass[a4paper]{article}

\usepackage{INTERSPEECH2021}
\usepackage{multirow}
\usepackage{subfigure}
\usepackage{wrapfig}
\usepackage{url}
\usepackage{color}
\usepackage{enumitem}

\title{AdaSpeech 3: Adaptive Text to Speech for Spontaneous Style}
\name{Yuzi Yan$^{\#1}$, Xu Tan$^{\$2*}$\thanks{$^*$Corresponding author: Xu Tan, xuta@microsoft.com}, Bohan Li$^{\%3}$, Guangyan Zhang$^{\&4}$, Tao Qin$^{\$5}$, Sheng Zhao$^{\%6}$,  Yuan Shen$^{\#7}$, \\Wei-Qiang Zhang$^{\#8}$, Tie-Yan Liu$^{\$9}$}
\address{$^\#$Department of Electronic Engineering, Tsinghua University, China\\ 
$^\$$Microsoft Research Asia, China \\ 
$^\%$Microsoft Azure Speech, China \\
$^\&$ Department of Electronic Engineering, The Chinese University of Hong Kong, China}
\email{$^1$yan-yz17@mails.tsinghua.edu.cn,\{$^{7}$shenyuan\_ee,$^{8}$wqzhang\}@tsinghua.edu.cn, \{$^2$xuta,$^3$bohli,$^5$taoqin,$^6$szhao,$^9$tyliu\}@microsoft.com, $^4$gyzhang@link.cuhk.edu.hk}

\begin{document}

\maketitle
\begin{abstract}
While recent text to speech (TTS) models perform very well in synthesizing reading-style (e.g., audiobook) speech, it is still challenging to synthesize spontaneous-style speech (e.g., podcast or conversation), mainly because of two reasons: 1) the lack of training data for spontaneous speech; 2) the difficulty in modeling the filled pauses (\emph{um} and \emph{uh}) and diverse rhythms in spontaneous speech. In this paper, we develop AdaSpeech 3, an adaptive TTS system that fine-tunes a well-trained reading-style TTS model for spontaneous-style speech. Specifically, 1) to insert filled pauses (FP) in the text sequence appropriately, we introduce an FP predictor to the TTS model; 2) to model the varying rhythms, we introduce a duration predictor based on mixture of experts (MoE), which contains three experts responsible for the generation of fast, medium and slow speech respectively, and fine-tune it as well as the pitch predictor for rhythm adaptation; 3) to adapt to other speaker timbre, we fine-tune some parameters in the decoder with few speech data. To address the challenge of lack of training data, we mine a spontaneous speech dataset to support our research this work and facilitate future research on spontaneous TTS. Experiments show that AdaSpeech 3 synthesizes speech with natural FP and rhythms in spontaneous styles, and achieves much better MOS and SMOS scores than previous adaptive TTS systems.

\end{abstract}


\section{Introduction}

Text to speech (TTS) aims to synthesize intelligible and natural speech~\cite{tan2021survey,wang2017tacotron,ping2018deep,ren2019fastspeech,ren2021fastspeech,yamagishi2007average}. While many models have been designed~\cite{wang2017tacotron,ren2019fastspeech} to synthesize reading-style speech and good quality has been achieved, the synthesis of spontaneous style, which usually occurs in scenarios like conversation, talk show or podcast~\cite{rochet2014take}, has not been well studied. According to previous studies~\cite{szekely2020breathing, winkworth1994variability,szekely2017synthesising,nagata2017dimensional}, spontaneous speech has two unique characteristics that distinguish it from reading-style speech: 1) The existence of filled pauses (FP) such as \emph{um} and \emph{uh}, which can increase the authenticity of perceived speakers~\cite{szekely2019spontaneous}. 2) Diverse rhythms. Spontaneous speech is more diverse in rhythms (which can be characterized by speaking rate and pitch, etc.)~\cite{wester2016evaluating}. Speakers sometimes lengthen or shorten a syllable and change their intonation in spontaneous scenarios~\cite{szekely2019casting}.

While there have been some previous works on building spontaneous TTS system, many of them are restricted to statistical parametric speech synthesis~\cite{dall2017statistical,sundaram2003empirical}. Although some works apply existing neural network based TTS system to synthesize spontaneous speech~\cite{szekely2019spontaneous}, the essential characteristics like FP and diverse rhythms have not been considered and modeled. That is, there are few neural network structures specifically designed for spontaneous speech synthesis. Besides, compared with reading-style speech, the training data for spontaneous speech is very limited, which hinders the research related. 


In this paper, we develop AdaSpeech 3, an adaptive TTS system that fine-tunes a well-trained reading-style TTS model for spontaneous-style speech\footnote{Considering spontaneous speech data is limited, adapting a TTS model trained on reading-style is a natural and good choice.}. AdaSpeech 3 is built upon a previous adaptive TTS system AdaSpeech~\cite{chen2021adaspeech}, with the following specific designs for spontaneous speech. 1) To insert filled pauses (FP) into the text sequence appropriately, we introduce an FP predictor to the TTS model. 2) Considering the difficulty in capturing the diverse rhythms, we design a duration predictor based on mixture of expert (MoE) and fine-tune it as well as the pitch predictor in the TTS model. To support the training of AdaSpeech 3, we mine a spontaneous speech dataset from a podcast collected from Internet, which contains three subsets that correspond to the key adaptation components in AdaSpeech 3: 1) SPON-FP, which consists of text-FP data pairs to train the PF predictor; 2) SPON-RHYTHM, which consists of text-pitch/duration data pairs to fine-tune the pitch predictor and MoE based duration predictor; 3) SPON-TIMBRE, which consists of text-speech data pairs for speaker timbre adaptation. 


AdaSpeech 3 has three main advantages. 1) Data efficiency. We adapt the model from a source TTS model that is well-trained on reading-style data, which is beneficial considering the shortage of spontaneous data. On the other hand, we only need the easily accessible text-FP and text-pitch/duration data pairs to adapt the FP/pitch/duration predictors, which reduces the data requirement for the scarce  high-quality spontaneous speech data. 2) Controllability. We can easily control the intensity of FP (how many FP occurs in an utterance) by adjusting the prediction threshold in FP predictor. 3) Transferability. We can transfer the spontaneous style to other speaker with few adaptation speech data no matter it is spontaneous or not.

We conduct several experiments to evaluate the performance of AdaSpeech 3\footnote{Synthesized speech samples can be found at \url{https://speechresearch.github.io/adaspeech3/}}. For voice quality, the MOS of AdaSpeech 3 is 0.24 higher than the baseline AdaSpeech~\cite{chen2021adaspeech} (a customized TTS model adapted by only the SPON-TIMBRE subset), in terms of different aspects such as naturalness, speaking rate, etc. For the similarity to spontaneous style, the SMOS of AdaSpeech 3 is 0.3 higher than that of AdaSpeech. Further method analyses demonstrate the effectiveness of each design in AdaSpeech 3. 

\section{Spontaneous Dataset Mining}

Since there is no public available spontaneous speech dataset, in this paper, we mine the dataset by ourself. 
Our dataset mining consists of several steps:

\begin{itemize}[leftmargin=*]
\item Data crawling. We collected the untranscribed spontaneous audio data from a podcast named “ThinkComputers”~\cite{szekely2019spontaneous} on Internet Archive (archive.org). We selected the data for the first 30 episodes with a total length of about 28 hours. 
\item ASR Transcription. We used an internal ASR tool to get the audio transcriptions, which include marked FP and the aligned timestamps. We used the given timestamps to cut the audio and text into fragments ranging from 7 to 10 second. We then convert each text sequence into phoneme sequence with grapheme-to-phoneme conversion~\cite{sun2019token}. 
\item SPON-FP dataset construction. To get the text-FP data pairs (denoted as SPON-FP) for the training of FP predictor, we remove the FP from the original phoneme sequence to get the sequence without FP, and extract the FP tag for each phoneme. The FP tag is defined as follows: If a phoneme is followed by an \emph{uh} or an \emph{um}, its corresponding tag is $1$ or $2$ respectively, otherwise its tag is $0$. Table~\ref{tab:FP} shows an example of text-FP data pair that consists of a phoneme sequence and its corresponding FP tags. 
\item SPON-RHYTHM dataset construction. We extract pitch from the spontaneous speech and get the duration through a forced-alignment tool~\cite{mcauliffe2017montreal}. The SPON-RHYTHM dataset contains the pitch and duration along with the phoneme sequence, which is used to fine-tune the pitch and duration predictors.
\item SPON-TIMBRE dataset construction. We use the phoneme sequence as well as the spontaneous speech data pairs to construct the SPON-TIMBRE dataset, which is used for speaker adaptation. 
\end{itemize}

The statistics of the three datasets: SPON-FP, SPON-RHYTHM and SPON-TIMBRE are shown in Table~\ref{tab:dataset}. For SPON-FP, there are 338 \emph{ums} and 2614 \emph{uhs} in total.

\vspace{-3pt}

\begin{table}[h]

\caption{An example of text-FP data pair. Phoneme w/o FP is the phoneme sequence with filled pauses (FP) removed.}
     \label{tab:FP}
     \centering
     \begin{tabular}{l | l}
     \toprule
     Raw text & \emph{It's called \textbf{um} right \textbf{uh} apple} \\
     Raw phomeme & \emph{ih t s k ao l \textcolor{red}{d} \textbf{ah m} r ay \textcolor{red}{t} \textbf{ah} ae p ax l} \\
     Phomeme w/o FP & \emph{ih t s k ao l \textcolor{red}{d} r ay \textcolor{red}{t} ae p ax l} \\
     FP tag & $0,0,0,0,0,0,\textcolor{red}{2},0,0,\textcolor{red}{1},0,0,0,0$ \\
     \bottomrule
    \end{tabular}
\end{table}

\vspace{-6pt}

\begin{table}[h]
  \caption{The statistics of the three datasets mined for spontaneous speech.}
  \label{tab:dataset}
  \centering
  \begin{tabular}{l  l  c}
    \toprule
    \textbf{Adaptation Step} &
    \textbf{Name} & 
    \textbf{Sentence Num} \\
    \midrule
    FP prediction & SPON-FP & 2952     \\
    Rhythm fine-tuning & SPON-RHYTHM & 14273    \\
    Speaker adaptation & SPON-TIMBRE & 50   \\
    \bottomrule
  \end{tabular}
\end{table}

\section{Method}

\begin{figure}[th]
  \centering
  \includegraphics[width=0.75\linewidth]{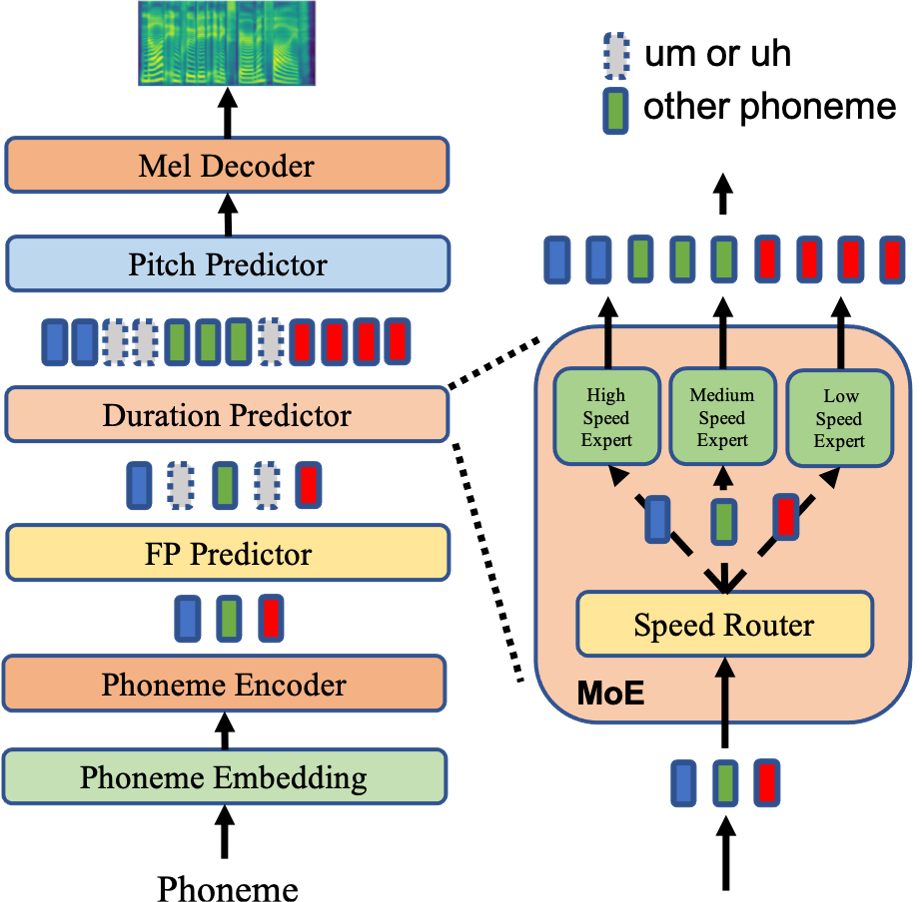}
  \caption{Model structure of AdaSpeech 3. The MoE based duration predictor is illustrated on the right side. Acoustic condition modeling and conditional normalization in AdaSpeech are also used but do not showed here mainly for simplicity.}
  \label{fig:arch}
\end{figure}

In this section, we first give an overview of our method, and then describe the important parts in the method in detail in the following subsections.

\subsection{Model Overview}

As shown in Figure~\ref{fig:arch}, AdaSpeech 3 is composed of: 1) a multi-speaker TTS model, which is based on the backbone of AdaSpeech~\cite{chen2021adaspeech}, a non-autoregressive adaptive TTS system with fast and high-quality speech synthesis as well as powerful and efficient adaptation capability; 2) an additional FP predictor, which is used to insert FP in the appropriate places in the sentences; 3) a newly designed duration predictor based on mixture of experts (MoE)~\cite{yuksel2012twenty,masoudnia2014mixture,avnimelech1999boosted}. In order to adapt the TTS model with spontaneous style, the pipeline of AdaSpeech 3 consists of the following steps: 

\begin{itemize}[leftmargin=*]
\item Source model training. Our source multi-speaker TTS model follows the structure used in AdaSpeech~\cite{chen2021adaspeech}. The phoneme encoder and the mel-spectrogram decoder both consist of 4 feed-forward Transformer blocks~\cite{vaswani2017attention,ren2019fastspeech,ren2021fastspeech}. The ground-truth duration is extracted from a forced-alignment tool~\cite{mcauliffe2017montreal}, and is used to train the duration predictor, which expands the hidden encoder output sequence to match to the length of mel-spectrogram sequence. The ground-truth durations are used for expansion in training while the predicted durations are used in inference. Similarly, the ground-truth pitch is used to train the pitch predictor and taken as input to the decoder in training while predicted pitch is used in inference. For the acoustic condition modeling and conditional layer normalization as used in ~\cite{chen2021adaspeech,yan2021adaspeech}, we integrate it in the phoneme encoder and mel decoder respectively, and do not explicitly show in Figure~\ref{fig:arch}.

\item FP predictor adaptation. FP is an important characteristic of spontaneous speech, and adding appropriate FP in a sentence can increase the spontaneous style. We add an FP predictor upon the phoneme encoder, which predicts the FP tag ($0$ for no FP, $1$ for \emph{uh} and $2$ for \emph{um}) for each phoneme, and insert the embedding of \emph{uh} or \emph{um} right after its corresponding phoneme hidden if FP tag is $1$ or $2$. SPON-FP is used for FP predictor adaptation. We will introduce the details in Section~\ref{sec:FP}.

\item Rhythm adaptation. In addition to FP, rhythm is another important factor that distinguishes spontaneous speech from reading-style speech. We adapt the rhythm (e.g., pitch and duration) to make it closer to spontaneous speech. We just adapt the pitch predictor normally with fine-tuning. However, in order to cover the large variations of duration in spontaneous speech, we design a mixture-of-expert (MoE) based duration predictor. SPON-RHYTHM is used for pitch and duration adaptation. We will introduce the details in Section~\ref{sec:Rhythm}.

\item Speaker timbre adaptation. We fine-tune the conditional layer normalization and speaker embedding as used in AdaSpeech~\cite{chen2021adaspeech} to adapt the timbre using adaptation data (e.g., the SPON-TIMBRE dataset or VCTK). In this way, we can transfer the spontaneous style to any custom voice with few adaptation data (e.g., 20 utterances), where the adaptation data is not necessarily to be spontaneous.
\end{itemize}

\subsection{FP Predictor Adaptation}
\label{sec:FP}

We build the FP predictor upon the phoneme encoder, which takes the phoneme hidden sequence as input and predicts the FP tags for each phoneme (a 3-class classification: $0$ for no FP, $1$ for \emph{uh}, and $2$ for \emph{um}). The FP predictor consists of 1) a 2-layer 1D-convolutional network with ReLU activation, each followed by the layer normalization and the dropout layer; 2) an extra linear layer and a 3-way softmax layer to predict probability of each tag. 

A challenge in training FP predictor is that the positive labels (i.e., 1 or 2) are extremely sparse. In other word, the ratio of sentences that contain FP is small and the ratio of tokens that contain positive FP tags (1 or 2) is extremely small (e.g., there are usually one \emph{um} or \emph{uh} in a sentence with more than 20 tokens).  In this way, the model will easily predict no FP (tag is 0) for all tokens. To alleviate the data sparse problem, we 1) use SPON-FP dataset mentioned in Table~\ref{tab:FP}, where sentences without FP are discarded to increase the density of positive labels, and 2) use the weighted cross entropy function as loss function: $L=- y_0\log{s_0}- \sigma \sum\limits_{i=1}^{2} y_i\log{s_i}$, 
where $[s_0,s_1,s_2]$ is the probability of the output belonging to three specific categories: $s_0$ for no FP, $s_1$ for \emph{uh} and $s_2$ for \emph{um}. $[y_0,y_1,y_2]$ represents the one-hot encoding of the ground-truth label accordingly. We can adjust $\sigma$ to ensure the data balance in the training. Besides, we can adjust the prediction threshold in FP predictor to control the intensity of FP (how many FP occurs in an utterance) in inference, which is described in detail in Section~\ref{sec:perFP}. After FP prediction, we can add the FP embeddings\footnote{Note that the FP embeddings added here are not shared with the phoneme embeddings in the input of encoder. } to the corresponding place in the phoneme hidden sequence as shown in Figure~\ref{fig:arch}. 


\subsection{Rhythm Adaptation}
\label{sec:Rhythm}

Duration and pitch are two main characteristics to determine the speech rhythm. To better adapt the rhythm of the TTS model, we analyze the difference in the distribution of pitch and duration between the spontaneous-style (SPON-RHYTHM) and reading-style (LibriTTS and VCTK) speech corpora. We have several observations: 1) For each speaker, pitch follows similar Gaussian distributions no matter it is from spontaneous or reading style; 2) However, as we mentioned before, speakers in spontaneous style sometimes lengthen or shorten a syllable, which causes the difference in duration distribution. The durations of each phoneme in LibriTTS and VCTK are mostly distributed between 0-25, while quite a few in SPON-RHYTHM is above 25. What is more, the duration distribution is more evenly distributed in the range of 0-40 (unlike the tailed distribution from LibriTTS and VCTK), which is caused by the speaker lengthening or shortening the phoneme in spontaneous style. The statistical results encourage us to enhance the duration predictor with more fine-grained speech rate control.

Thus, we adopt a simple strategy (just fine-tuning) for pitch adaptation and make a targeted design for duration adaptation. Specifically, we introduce mixture of experts (MoE) in the duration predictor to better capture the large duration range, where the MoE consists of a speed router and three expert predictors. We describe the process to build the MoE based duration predictor as follows: 1) We categorize the phoneme durations evenly into low, medium and high speed (corresponding to large, medium and small durations), which we call speed tag. 2) We use the phoneme hidden and its corresponding speed tag to train a speed router, which shares the same structure with the FP predictor. 3) We initialize the three duration experts from the duration predictor in the well-trained source TTS model, where each duration expert is fine-tuned with the phoneme hidden and duration pairs in low, medium and high speed respectively, as shown in Figure~\ref{fig:arch}.  4) The ground-truth duration is used to expand the phoneme hidden sequence in training, while in inference, the predicted duration is used. To get the predicted duration from MoE, we calculate a weighted average of the predictions from three experts, where the weights are the probabilities predicted by the speed router. 


\section{Experiments and Results}

\subsection{Experimental Setting}
We use LibriTTS~\cite{zen2019libritts} dataset (reading-style) to train the source TTS model, and use SPON-FP, SPON-RHYTHM and SPON-TIMBRE to adapt the FP predictor, duration/pitch predictor and speaker timbre respectively. To evaluate the ability to transfer spontaneous style to reading-style speaker, we randomly select two males and two females (each with 50 utterances) in VCTK~\cite{veaux2016superseded} (reading-style) for speaker adaptation.

The hidden dimension (including the embedding size, the hidden size in self-attention, and the input and output size of feed-forward network) is set to 256. The attention head, the feed-forward filter size and kernel size are set to 2, 1024 and 9 respectively. The output linear layer converts the 256-dimensional hidden into 80-dimensional mel-spectrogram.

We first take 100,000 steps to train the source TTS model on 4 NVIDIA P40 GPUs, and then takes 4,000 steps to adapt the FP predictor, and then takes 4,000 steps to adapt the MoE based duration predictor (including the speed router and the three experts) and the pitch predictor. We further take 2,000 steps following the setting used in~\cite{chen2021adaspeech} for speaker adaptation. All the adaptations are conducted in 1 NVIDIA P40 GPU. 
The Adam optimizer is used with $\beta_1=0.9$, $\beta_2=0.98$, $\epsilon=10^{-9}$. The quality of the synthesized speech is evaluated with MOS (Mean Opinion Score)~\cite{viswanathan2005measuring}, SMOS (Similarity MOS) and CMOS (comparison MOS).

\begin{table}[th]
  \caption{The SMOS results of AdaSpeech 3 and the baseline.}
  \label{tab1}
  \centering
  \begin{tabular}{l  c}
    \toprule
    \multicolumn{1}{c}{\textbf{Setting}} & 
    \multicolumn{1}{c}{\textbf{SMOS}} \\
    \midrule
    GT              &  $4.33\pm 0.14$   \\
    GT mel+Vocoder  &  $4.07\pm 0.14$  \\
    AdaSpeech       &  $3.45\pm 0.18$   \\
    \midrule
    AdaSpeech 3      &  $3.75\pm 0.16$     \\
    \bottomrule
  \end{tabular}
\end{table}

\begin{table}[th]
  \caption{The MOS results of AdaSpeech 3 and the baseline in terms of different evaluation aspects.}
  \label{tab:mos}
  \centering
  \begin{tabular}{ l   c c c}
    \toprule
    \textbf{Setting} & 
    Naturalness &
    Pause &
    Speaking Rate\\
    \midrule
     GT             &  $4.14\pm 0.06$    &  $4.01\pm 0.06$ &  $3.04\pm 0.06$     \\
     GT mel+Voc &  $3.84\pm 0.06$   &  $3.78\pm 0.06$ &  $3.06\pm 0.08$   \\
     AdaSpeech  &  $3.21\pm 0.06$  &  $3.36\pm 0.06$ &  $2.66\pm 0.08$   \\
     \midrule
     AdaSpeech 3 &  $3.45\pm 0.06$  &  $3.53\pm 0.06$ &  $2.79\pm 0.06$    \\
    \bottomrule
  \end{tabular}
\end{table}

\subsection{The Quality of AdaSpeech 3}

We compare the synthesized speech of AdaSpeech 3 with other settings, including: 1) GT, the ground truth recordings; 2) GT mel + Vocoder (Voc), using MelGAN vocoder~\cite{kumar2019melgan} to synthesize waveform from ground-truth mel-spectrogram; 3) AdaSpeech~\cite{chen2021adaspeech}, a previous reading-style TTS adaptation system. AdaSpeech 3 uses the same source TTS model as in AdaSpeech, but differs in that AdaSpeech 3 performs FP, rhythm and speaker timbre adaptation while AdaSpeech only performs speaker timbre adaptation. The comparison with AdaSpeech can verify the effectiveness of the specific adaptation designs in AdaSpeech 3 for spontaneous style\footnote{Synthesized speech samples can be found at \url{https://speechresearch.github.io/adaspeech3/}.}.

We ask 20 native English speakers to evaluate the similarity and naturalness of the synthesized speech (15 sentences) by each setting. We evaluate the similarity with spontaneous style in terms of pitch, speed, volume, intonation, rhythm and stress, and show the SMOS results in Table~\ref{tab1}. We also evaluate the quality of the synthesized speech in different aspects, including naturalness, inappropriate pause and speaking rate, and show the MOS results in Table~\ref{tab:mos}. It can be seen that AdaSpeech 3 outperforms the baseline AdaSpeech in all aspects. 

\subsection{Method Analyses}
\label{sec:perFP}

\noindent \textbf{Effectiveness of FP prediction.}
We compare our FP prediction with a setting that inserts the same amount of FP randomly in the sentence, and conduct a CMOS evaluation focusing on the rationality of FP insertion. 
The result shows that our FP prediction achieves $0.156$ CMOS higher than the random insertion setting, demonstrating the effectiveness of our FP prediction. 

As described in Section~\ref{sec:FP}, we can control the intensity of FP by changing the prediction threshold. For prediction probability $s_0, s_1, s_2$ which corresponds to the three categories (no FP, \emph{uh}, and \emph{um}). If $s_0$ is larger than a threshold $T$, we classify it as no FP, otherwise, we classify it to the FP with larger probability. The recall, precision and accuracy rate are  measured on 40 test sentences from SPON-FP. When $T$ increases from $0.10$ to $0.99$, the recall rate increases from $0.800$ to $0.950$ (at the cost of a slight drop in precision), which means the system has a higher tendency to insert FP. 


\begin{table}[th]
  \caption{The CMOS results in the ablation studies of rhythm fine-tuning. w/o MoE: fine-tuning with the single duration predictor instead of MoE; w/o duration adaptation: using the single duration predictor without fine-tuning. w/o pitch adaptation: using the pitch predictor without fine-tuning. w/o pitch/duration adaptation: using the pitch/duration predictor without fine-tuning.}
  \label{tab:aba1}
  \centering
  \begin{tabular}{l  c c}
    \toprule
    \textbf{Setting} & 
    \textbf{with FP} &
    \textbf{without FP} \\
    \midrule
    AdaSpeech 3   &  / & /  \\
    \midrule
    ~~w/o MoE &  $-0.332$  &  $-0.274$      \\
    ~~w/o duration adaptation &  $-0.356$ &$-0.289$     \\
    ~~w/o pitch adaptation   & $-0.157$  &  $-0.112$       \\
    ~~w/o pitch/duration adaptation & $-0.384$  & $-0.308$     \\
    \bottomrule
  \end{tabular}
\end{table}

\noindent \textbf{Effectiveness of rhythm adaptation.} We compare our default adaptation with different variations and conduct CMOS evaluation focusing on tone and rhythm. The results are evaluated in two settings (with and without FP insertion) and are shown in Table~\ref{tab:aba1}. We have several observations: 1) Replacing MoE based duration predictor with a single duration predictor but still with fine-tuning (denoted as w/o MoE) causes genral quality drop, which demonstrates the advantages of MoE design in duration predictor. 2) Further removing fine-tuning on the single duration predictor (denoted as w/o duration adaptation) causes more quality drop, which verifies the effectiveness of fine-tuning. 3) Removing pitch fine-tuning (denoted as w/o pitch adaptation) causes quality drop, but with smaller drops than removing duration adaptation. 4) Removing both pitch and duration adaptation (denoted as w/o pitch/duration adaptation) further degrades the quality. 5) The quality degradation in all variations are more severe in the ``with FP'' setting compared with the ``without FP'' setting, indicating that rhythm fine-tuning can improve the spontaneous style especially when FP is inserted, showing that rhythm fine-tuning and FP prediction are complementary for spontaneous speech synthesis.


\subsection{Cross-Speaker Spontaneous-Style Transfer}

AdaSpeech 3 can also transfer the spontaneous style to other speakers originally without spontaneous data, with only a few extra reading-style data for adaptation. We select two males and two females from VCTK for speaker timbre adaptation and synthesize speech based on the same text in the test set of SPON-TIMBRE. We evaluate the improvement in spontaneous style by conducting a CMOS test focusing on tone and rhythm. AdaSpeech 3 achieves $0.175$ CMOS score higher than AdaSpeech in male's voice and $0.205$ higher in female's voice.

\section{Conclusion}
In this paper, we develop AdaSpeech 3, an adaptive TTS system for spontaneous-style speech. To enable the research work in this paper and facilitate potential future research in this area, we mine a spontaneous speech dataset that contains three subsets (SPON-FP, SPON-RHYTHM and SPON-TIMBRE) for spontaneous-style adaptation. We design an FP predictor to insert filled pauses appropriately in the sentence and introduce a mixture-of-expert based duration predictor to capture the diverse rhythms in spontaneous speech. AdaSpeech 3 achieves better quality in the synthesized spontaneous speech in terms of MOS, SMOS and CMOS, compared with a strong adaptive TTS system AdaSpeech. For future work, we will explore more characteristics in spontaneous speech such as repetition and discourse marker to improve spontaneous TTS.

\bibliographystyle{IEEEtran}

\bibliography{mybib}

\begin{thebibliography}{10}
\providecommand{\url}[1]{#1}
\csname url@samestyle\endcsname
\providecommand{\newblock}{\relax}
\providecommand{\bibinfo}[2]{#2}
\providecommand{\BIBentrySTDinterwordspacing}{\spaceskip=0pt\relax}
\providecommand{\BIBentryALTinterwordstretchfactor}{4}
\providecommand{\BIBentryALTinterwordspacing}{\spaceskip=\fontdimen2\font plus
\BIBentryALTinterwordstretchfactor\fontdimen3\font minus
  \fontdimen4\font\relax}
\providecommand{\BIBforeignlanguage}[2]{{%
\expandafter\ifx\csname l@#1\endcsname\relax
\typeout{** WARNING: IEEEtran.bst: No hyphenation pattern has been}%
\typeout{** loaded for the language `#1'. Using the pattern for}%
\typeout{** the default language instead.}%
\else
\language=\csname l@#1\endcsname
\fi
#2}}
\providecommand{\BIBdecl}{\relax}
\BIBdecl

\bibitem{tan2021survey}
X.~Tan, T.~Qin, F.~Soong, and T.-Y. Liu, ``A survey on neural speech
  synthesis,'' \emph{arXiv preprint arXiv:2106.15561}, 2021.

\bibitem{wang2017tacotron}
Y.~Wang, R.~Skerry-Ryan, D.~Stanton, Y.~Wu, R.~J. Weiss, N.~Jaitly, Z.~Yang,
  Y.~Xiao, Z.~Chen, S.~Bengio \emph{et~al.}, ``Tacotron: Towards end-to-end
  speech synthesis,'' \emph{arXiv preprint arXiv:1703.10135}, 2017.

\bibitem{ping2018deep}
W.~Ping, K.~Peng, A.~Gibiansky, S.~O. Arik, A.~Kannan, S.~Narang, J.~Raiman,
  and J.~Miller, ``Deep voice 3: 2000-speaker neural text-to-speech,'' in
  \emph{International Conference on Learning Representations}, 2018.

\bibitem{ren2019fastspeech}
Y.~Ren, Y.~Ruan, X.~Tan, T.~Qin, S.~Zhao, Z.~Zhao, and T.-Y. Liu, ``Fastspeech:
  Fast, robust and controllable text to speech,'' in \emph{NIPS}, 2019, pp.
  3165--3174.

\bibitem{ren2021fastspeech}
\BIBentryALTinterwordspacing
Y.~Ren, C.~Hu, X.~Tan, T.~Qin, S.~Zhao, Z.~Zhao, and T.-Y. Liu, ``Fastspeech 2:
  Fast and high-quality end-to-end text to speech,'' in \emph{International
  Conference on Learning Representations}, 2021. [Online]. Available:
  \url{https://openreview.net/forum?id=piLPYqxtWuA}
\BIBentrySTDinterwordspacing

\bibitem{yamagishi2007average}
J.~Yamagishi and T.~Kobayashi, ``Average-voice-based speech synthesis using
  hsmm-based speaker adaptation and adaptive training,'' \emph{IEICE
  TRANSACTIONS on Information and Systems}, vol.~90, no.~2, pp. 533--543, 2007.

\bibitem{rochet2014take}
A.~Rochet-Capellan and S.~Fuchs, ``Take a breath and take the turn: how
  breathing meets turns in spontaneous dialogue,'' \emph{Philosophical
  Transactions of the Royal Society B: Biological Sciences}, vol. 369, no.
  1658, p. 20130399, 2014.

\bibitem{szekely2020breathing}
{\'E}.~Sz{\'e}kely, G.~E. Henter, J.~Beskow, and J.~Gustafson, ``Breathing and
  speech planning in spontaneous speech synthesis,'' in \emph{ICASSP 2020-2020
  IEEE International Conference on Acoustics, Speech and Signal Processing
  (ICASSP)}.\hskip 1em plus 0.5em minus 0.4em\relax IEEE, 2020, pp. 7649--7653.

\bibitem{winkworth1994variability}
A.~L. Winkworth, P.~J. Davis, E.~Ellis, and R.~D. Adams, ``Variability and
  consistency in speech breathing during reading: Lung volumes, speech
  intensity, and linguistic factors,'' \emph{Journal of Speech, Language, and
  Hearing Research}, vol.~37, no.~3, pp. 535--556, 1994.

\bibitem{szekely2017synthesising}
E.~Sz{\'e}kely, J.~Mendelson, and J.~Gustafson, ``Synthesising uncertainty: The
  interplay of vocal effort and hesitation disfluencies.'' in
  \emph{INTERSPEECH}, 2017, pp. 804--808.

\bibitem{nagata2017dimensional}
T.~Nagata, H.~Mori, and T.~Nose, ``Dimensional paralinguistic information
  control based on multiple-regression hsmm for spontaneous dialogue speech
  synthesis with robust parameter estimation,'' \emph{Speech Communication},
  vol.~88, pp. 137--148, 2017.

\bibitem{szekely2019spontaneous}
{\'E}.~Sz{\'e}kely, G.~E. Henter, J.~Beskow, and J.~Gustafson, ``Spontaneous
  conversational speech synthesis from found data,'' in \emph{Interspeech},
  2019.

\bibitem{wester2016evaluating}
M.~Wester, O.~Watts, and G.~E. Henter, ``Evaluating comprehension of natural
  and synthetic conversational speech,'' in \emph{Proc. Speech Prosody},
  vol.~8, 2016, pp. 736--740.

\bibitem{szekely2019casting}
{\'E}.~Sz{\'e}kely, G.~E. Henter, and J.~Gustafson, ``Casting to corpus:
  Segmenting and selecting spontaneous dialogue for tts with a cnn-lstm
  speaker-dependent breath detector,'' in \emph{ICASSP 2019-2019 IEEE
  International Conference on Acoustics, Speech and Signal Processing
  (ICASSP)}.\hskip 1em plus 0.5em minus 0.4em\relax IEEE, 2019, pp. 6925--6929.

\bibitem{dall2017statistical}
R.~Dall, ``Statistical parametric speech synthesis using conversational data
  and phenomena,'' 2017.

\bibitem{sundaram2003empirical}
S.~Sundaram and S.~Narayanan, ``An empirical text transformation method for
  spontaneous speech synthesizers,'' in \emph{Eighth European Conference on
  Speech Communication and Technology}, 2003.

\bibitem{chen2021adaspeech}
\BIBentryALTinterwordspacing
M.~Chen, X.~Tan, B.~Li, Y.~Liu, T.~Qin, S.~Zhao, and T.-Y. Liu, ``Adaspeech:
  Adaptive text to speech for custom voice,'' in \emph{International Conference
  on Learning Representations}, 2021. [Online]. Available:
  \url{https://openreview.net/forum?id=Drynvt7gg4L}
\BIBentrySTDinterwordspacing

\bibitem{sun2019token}
H.~Sun, X.~Tan, J.-W. Gan, H.~Liu, S.~Zhao, T.~Qin, and T.-Y. Liu,
  ``Token-level ensemble distillation for grapheme-to-phoneme conversion,'' in
  \emph{INTERSPEECH}, 2019.

\bibitem{mcauliffe2017montreal}
M.~McAuliffe, M.~Socolof, S.~Mihuc, M.~Wagner, and M.~Sonderegger, ``Montreal
  forced aligner: Trainable text-speech alignment using kaldi.'' in
  \emph{Interspeech}, 2017, pp. 498--502.

\bibitem{yuksel2012twenty}
S.~E. Yuksel, J.~N. Wilson, and P.~D. Gader, ``Twenty years of mixture of
  experts,'' \emph{IEEE transactions on neural networks and learning systems},
  vol.~23, no.~8, pp. 1177--1193, 2012.

\bibitem{masoudnia2014mixture}
S.~Masoudnia and R.~Ebrahimpour, ``Mixture of experts: a literature survey,''
  \emph{Artificial Intelligence Review}, vol.~42, no.~2, pp. 275--293, 2014.

\bibitem{avnimelech1999boosted}
R.~Avnimelech and N.~Intrator, ``Boosted mixture of experts: An ensemble
  learning scheme,'' \emph{Neural computation}, vol.~11, no.~2, pp. 483--497,
  1999.

\bibitem{vaswani2017attention}
A.~Vaswani, N.~Shazeer, N.~Parmar, J.~Uszkoreit, L.~Jones, A.~N. Gomez,
  {\L}.~Kaiser, and I.~Polosukhin, ``Attention is all you need,'' in
  \emph{NIPS}, 2017, pp. 5998--6008.

\bibitem{yan2021adaspeech}
Y.~Yan, X.~Tan, B.~Li, T.~Qin, S.~Zhao, Y.~Shen, and T.-Y. Liu, ``Adaspeech 2:
  Adaptive text to speech with untranscribed data,'' in \emph{ICASSP 2021-2021
  IEEE International Conference on Acoustics, Speech and Signal Processing
  (ICASSP)}.\hskip 1em plus 0.5em minus 0.4em\relax IEEE, 2021, pp. 6613--6617.

\bibitem{zen2019libritts}
H.~Zen, V.~Dang, R.~Clark, Y.~Zhang, R.~J. Weiss, Y.~Jia, Z.~Chen, and Y.~Wu,
  ``Libritts: A corpus derived from librispeech for text-to-speech,''
  \emph{arXiv preprint arXiv:1904.02882}, 2019.

\bibitem{veaux2016superseded}
C.~Veaux, J.~Yamagishi, K.~MacDonald \emph{et~al.}, ``Superseded-cstr vctk
  corpus: English multi-speaker corpus for cstr voice cloning toolkit,'' 2016.

\bibitem{viswanathan2005measuring}
M.~Viswanathan and M.~Viswanathan, ``Measuring speech quality for
  text-to-speech systems: development and assessment of a modified mean opinion
  score (mos) scale,'' \emph{Computer Speech \& Language}, vol.~19, no.~1, pp.
  55--83, 2005.

\bibitem{kumar2019melgan}
K.~Kumar, R.~Kumar, T.~de~Boissiere, L.~Gestin, W.~Z. Teoh, J.~Sotelo,
  A.~de~Br{\'e}bisson, Y.~Bengio, and A.~C. Courville, ``Melgan: Generative
  adversarial networks for conditional waveform synthesis,'' in \emph{NIPS},
  2019, pp. 14\,910--14\,921.

\end{thebibliography}


\end{document}